\documentclass[journal]{IEEEtran}
\usepackage{booktabs}

\usepackage{makecell}
\newcommand{\tabincell}[2]{\begin{tabular}{@{}#1@{}}#2\end{tabular}}

\usepackage{xcolor}

\usepackage{csquotes}
\MakeOuterQuote{"}

\usepackage{graphicx}

\usepackage{url}
\begin{document}
\bstctlcite{IEEEexample:BSTcontrol}
%
\title{Exploring Generative AI Techniques \\in Government: A Case Study}
%
%
%

\author{Sunyi Liu$^*$,
        Mengzhe Geng$^{*}$,
        and~Rebecca Hart
    \thanks{$^*$Equal contribution.\\
        \indent Sunyi Liu is with University of Toronto, Toronto, Canada. Work done during an internship at the National Research Council of Canada (email: alysa.liu@mail.utoronto.ca).  \\
        \indent Mengzhe Geng and Rebecca Hart are with the National Research Council of Canada, Ottawa, Canada (e-mail: \{Mengzhe.Geng,Rebecca.Hart\}@nrc-cnrc.gc.ca). \\
        \indent Mengzhe Geng is the corresponding author.}
}

%
%

\markboth{Journal of \LaTeX\ Class Files,~Vol.~14, No.~8, August~2015}%
{Shell \MakeLowercase{\textit{et al.}}: Bare Demo of IEEEtran.cls for IEEE Journals}
%



\maketitle

\begin{abstract}
The swift progress of Generative Artificial intelligence (GenAI), notably Large Language Models (LLMs), is reshaping the digital landscape. Recognizing this transformative potential, the National Research Council of Canada (NRC) launched a pilot initiative to explore the integration of GenAI techniques into its daily operation for performance excellence, where 22 projects were launched in May 2024. Within these projects, this paper presents the development of the intelligent agent \textit{Pubbie} as a case study, targeting the automation of performance measurement, data management and insight reporting at the NRC. Cutting-edge techniques are explored, including LLM orchestration and semantic embedding via RoBERTa, while strategic fine-tuning and few-shot learning approaches are incorporated to infuse domain knowledge at an affordable cost. The user-friendly interface of \textit{Pubbie} allows general government users to input queries in natural language and easily upload or download files with a simple button click, greatly reducing manual efforts and accessibility barriers.
\end{abstract}


%
\IEEEpeerreviewmaketitle

\section{Introduction}
\label{sec:intro}
 
\IEEEPARstart{O}{ver} the past few years, the rapid progress of Generative Artificial Intelligence (GenAI)~\cite{feuerriegel2024generative}, particularly Large Language Models (LLMs)~\cite{chang2024survey}, marked the beginning of a new technological era. From chatbots automatically responding to customer queries~\cite{kolasani2023optimizing} to toolkits facilitating automated content creation~\cite{yang2024harnessing}, GenAI is reshaping the digital landscape. The ability of LLMs to generate context-aware responses~\cite{yang2024harnessing} and their strong reasoning abilities~\cite{sun2023survey} endow them with considerable potential for applications across a spectrum of industries. 

Recognizing this transformative potential, the National Research Council of Canada (NRC) has launched a program to explore the integration of GenAI into its operational framework and corporate mechanisms. As a pilot initiative, the NRC called for proposals from its corporate divisions to identify processes that could leverage the capabilities of GenAI and LLMs. After assessing 34 proposals for technical feasibility, they were grouped into 6 use cases focusing on enhancing the efficiency and consistency of various operations such as data collection, criteria evaluation, answer synthesis, request triaging, project scoping, and data reporting. In May 2024, 22 projects were launched. A cohort of 22 students with backgrounds in AI was recruited from Canadian universities to explore GenAI-driven solutions for each project, co-supervised by a non-technical business process owner and a researcher with a computer science background.

This paper delves into one project within this pilot program as a case study, aiming to improve and automate the workflows of performance measurement, data management and insight reporting at the NRC. As a federal agency, the NRC collects extensive data to fulfill accountability requirements~\cite{tbcs2016} and presently relies heavily on manual efforts for data collection, validation, and reporting. In the case study, we propose to leverage GenAI to streamline such processes, minimize manual workload, and boost the agility and efficiency of reporting.

A key obstacle with the NRC’s current reporting approach involves associating publications produced by NRC researchers each year with the NRC's collaborative challenge programs\footnote{\url{https://nrc.canada.ca/en/research-development/research-collaboration/programs/challenge-programs}} that support the research. In light of the missing linkage between publications and the NRC's challenge programs in publicly accessible databases like Scopus~\cite{burnham2006scopus}, our current approach entails separate tracking processes and manually matching publications with the relevant challenge program. With NRC researchers producing over 1,000 articles yearly, such a process is both time-consuming and error-prone. Furthermore, while NRC officers have access to the organization's publication data, extracting insights from this dataset involves labor-intensive manual searches through quarterly Excel files.

While conventional AI models might appear capable of streamlining the aforementioned matching process, there are two major challenges. Firstly, the team at the NRC responsible for performance measurement has limited computer science knowledge, making conventional AI models, which typically require familiarity with command-line operations~\cite{shotts2019linux}, unsuitable. Secondly, the diverse formats of inquiries that may arise for reporting purposes present significant challenges for traditional AI models structured to manage fixed-format inputs and yield fixed-format outputs. On the other hand, as linking publications with the NRC's challenge programs is a specialized task, off-the-shelf GenAI products lack the relevant domain knowledge to handle it properly. 

To address these issues, this paper introduces an intelligent agent named \textit{Pubbie}, leveraging the generalization and reasoning abilities of GenAI and LLMs. \textit{Pubbie} has the capacity to automatically link publications with the relevant NRC challenge program and respond to open-ended user inquiries. Our proposed methodology follows a three-stage approach that incorporates a backend SQLite database~\cite{kreibich2010using},  a fine-tuned RoBERTa model~\cite{zhuang2021obustly}, and LLM orchestration~\cite{rasal2024navigating} via Semantic Kernel~\cite{microsoft_semantic_kernel}:

\begin{figure*}[!ht]
  \centering
  \includegraphics[scale=0.56]{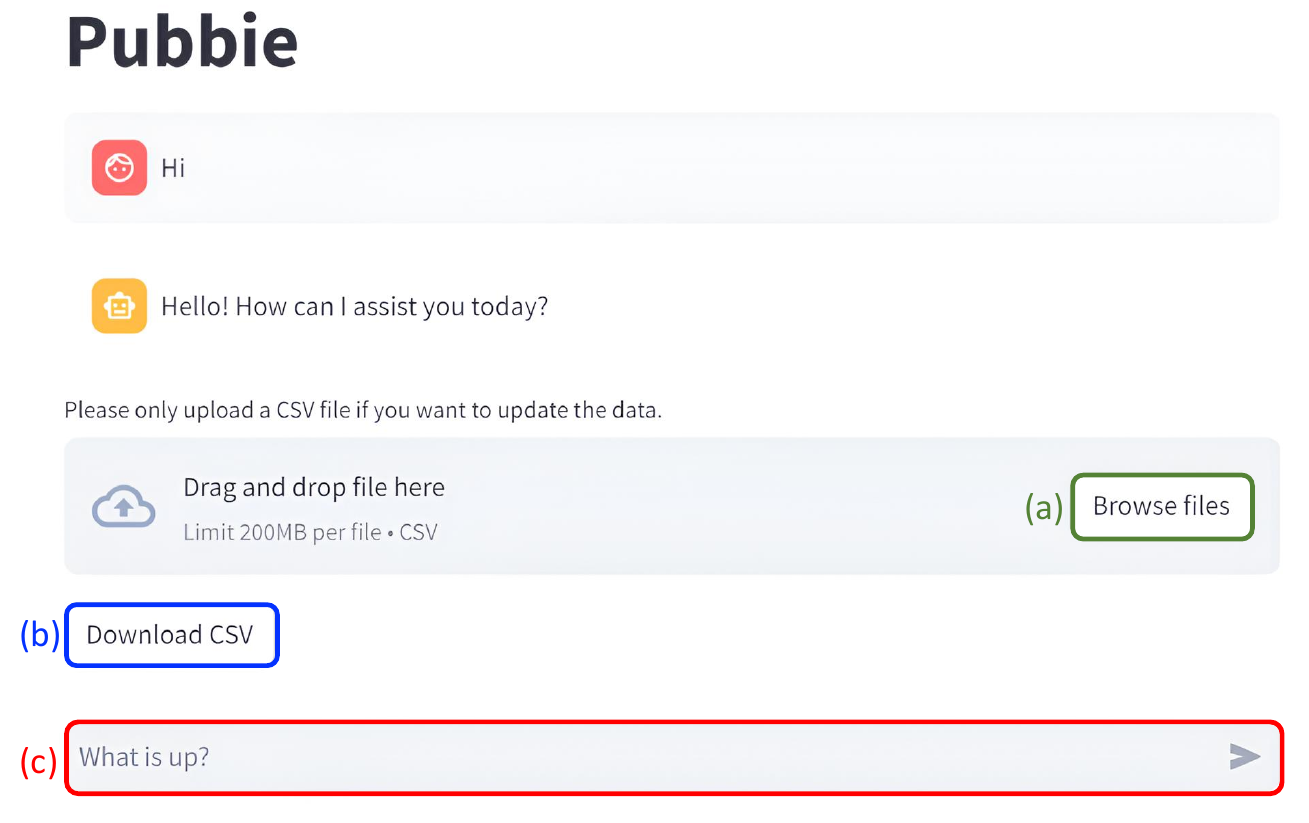}
  \caption{The interface of our intelligent agent, \textit{Pubbie}, including (a) a button for uploading the CSV file; (b) a button for downloading the CSV file; and (c) a chat box for inquiries in natural language.}
  \label{fig:interface}
\end{figure*}

\vspace{0.2cm}
\begin{itemize}
    \item \textbf{Stage 1:} Fine-tune a RoBERTa model on a manually labeled dataset to predict challenge programs for new publications.

    \item \textbf{Stage 2:} Incorporate prompt templates and domain data into the Semantic Kernel framework for few-shot learning. The fine-tuned LLM is then to generate SQL queries for creating a database containing publications and the affiliated programs.
    
    \item \textbf{Stage 3:} Establish a three-fold workflow for LLM orchestration to incorporate historical context, categorize user inputs, allocate LLMs to handle them accordingly, and then generate context-aware responses.
\end{itemize}
\vspace{0.2cm}

We further utilize Streamlit~\cite{khorasani2022web} to construct a user-friendly interface for \textit{Pubbie}. As depicted in Fig.~\ref{fig:interface}, users can easily type their queries in the chat box (circled in red) and upload (circled in green) or download (circled in blue) the file with a simple button click, which greatly reduces the accessibility barriers.

The main contributions of this paper are summarized below: 

\begin{enumerate}
    \item \textbf{First exploration of GenAI techniques in government operations:} To the best of our knowledge, this paper presents the first exploration to leverage GenAI and LLMs to empower the daily operation of government for operational excellence. The NRC’s GenAI initiative encompasses 22 projects spanning six key areas, including data, criteria evaluation, answer synthesis, request triaging, project scoping, and data reporting. Within the initiative, we provide an in-depth case study of one project concentrating on automating the workflow of performance measurement, data management and insight reporting at the NRC.
    
    \item \textbf{User-friendly and cost-affordable intelligent agent:} Tailored for general users in the government, the proposed intelligent agent, \textit{Pubbie}, eradicates the prerequisite for a computer science background to leverage cutting-edge technologies for operational excellence. The user-friendly design and interface of our agent extend its applicability and adaptability to diverse scenarios, all while ensuring that the costs and resources for development are affordable for a general government department without access to high-performance computing (HPC) facilities.
    
    \item \textbf{Customized LLMs for domain-aware and context-aware responses:} Instead of relying on out-of-the-box LLMs and embedding models as they are, we develop customized LLMs by conducting strategic fine-tuning or few-shot learning in our task domain, tailoring them to provide domain-aware and context-aware responses. To further address inherent limitations of LLMs such as constraints on input length, we innovatively integrate a classification model and an SQLite database into the agent, enabling collaborative efforts of conventional and generative AI techniques.
\end{enumerate}


 

\section{Methodology}
\label{sec:method}
 
This section presents the key techniques we leverage to develop the intelligent agent \textit{Pubbie}, including LLM orchestration~\cite{rasal2024navigating} via Semantic Kernel~\cite{microsoft_semantic_kernel} for integration, BERT~\cite{devlin2019bert} and RoBERTa~\cite{zhuang2021obustly} models for extracting semantic embeddings, as well as the SQLite database~\cite{kreibich2010using} for data storage. 

\subsection{LLM Orchestration}
Large Language Model (LLM) orchestration entails integrating and coordinating multiple LLMs or non-LLM models to boost performance~\cite{rasal2024navigating}.  Instead of tasking one LLM with addressing the entire problem, this approach involves decomposing the user's query into concrete sub-problems. These sub-problems are then addressed in parallel by carefully selected LLMs or non-LLM models, while the orchestrating framework oversees the process and consolidates the solutions into one comprehensive final response. We implement LLM orchestration using the Semantic Kernel~\cite{microsoft_semantic_kernel} framework provided by Azure OpenAI, which provides affordable API access to cutting-edge LLMs developed by OpenAI and deployed on the Azure cloud platform. 

\subsection{BERT and RoBERTa}
BERT (Bidirectional Encoder Representations from Transformers)~\cite{devlin2019bert} is widely recognized in the literature for its exceptional ability to derive bidirectional context-aware deep representations from unlabeled text. It facilitates straightforward fine-tuning of the pre-trained model on a diverse array of downstream tasks, such as text classification and semantic analysis. 

Building upon the success of BERT, RoBERTa (Robustly Optimized BERT Pretraining Approach)~\cite{zhuang2021obustly} further optimizes the pretraining process by refining masking strategies \& training data size and tuning hyperparameters. The resulting pre-trained model surpasses the pre-trained BERT model on a variety of downstream tasks~\cite{devlin2019bert}.

\subsection{SQLite Database}
SQLite~\cite{kreibich2010using} is a lightweight, self-contained, and zero-configuration database engine, eliminating the need for an external server process. By housing the entire database within a single file on a local disk, it bolsters portability and effortless integration across various applications. Its serverless nature enables keeping sensitive information in a closely controlled environment, rendering it an ideal choice for governmental users handling confidential information~\cite{lloyd2024canada}.

\section{The Intelligent Agent Pubbie: A Case Study}
\label{sec:case}

This section provides an in-depth exploration of the development of our intelligent agent \textit{Pubbie}, spotlighting it as a prime case study within the GenAI initiative by the National Research Council of Canada (NRC).

\subsection{Background}
\label{sec:case_background}


The NRC gathers substantial performance data quarterly, structured primarily to address information needs identified at the time of data collection. However, challenges arise when attempting to repurpose this data to answer new, related queries that emerge at a later stage. One notable example is the task of accurately associating individual publications with the NRC’s challenge programs. In such cases, the performance team currently relies on manual processes to manipulate and adapt the existing data, including: 
\begin{itemize}
    \item \textbf{Performing Excel-based searches}: Identifying relevant publications through a series of searches using various related search terms.  
    \item \textbf{Manual review of publication lists}: Reviewing the complete list of publications and tagging relevant ones, often requiring subject matter expertise to identify specific topics or domains. 
    \item \textbf{Maintaining separate lists for challenge programs}: Creating and updating a separate list of publications produced by the team, which are then cross-validated against the complete publication dataset. 
\end{itemize}
The NRC's publication dataset is sourced from the publicly accessible Scopus database~\cite{burnham2006scopus} and organized in CSV format, with 20 attributes such as titles, publication years, authors, affiliations, keywords, and Scopus EIDs. It includes 6,567 NRC publications, of which 656 are manually labeled with their corresponding challenge programs\footnote{According to the NRC's challenge program structure, a publication may be linked to a program or remain unaffiliated.}. The proposed intelligent agent, \textit{Pubbie}, aims to streamline the association of publications with challenge programs and query handling.

\subsection{Tasks and Challenges}
Given the specific requirements of the NRC's performance measurement process, our intelligent agent \textit{Pubbie} is designed to handle three main tasks: 1) addressing open-ended inquiries about the NRC's publications, including their affiliated challenge programs; 2) updating the publication database when new information is available; and 3) exporting details of retrieved publications to a CSV file for easy access and sharing. These tasks lead to several challenges we need to tackle:
\begin{itemize}
    \item As highlighted in Section~\ref{sec:intro}, relying solely on traditional AI models is unfeasible due to their inability to handle flexible queries, while out-of-the-box LLM solutions are incapable of generating accurate domain-aware responses; \cite{lu2024}

    \item Inputting the entire dataset into an LLM in one go is impossible due to input length constraints, but this is necessary to address queries requiring aggregated dataset information;
    \item The agent is expected to respond in time while operating using limited computing resources available to a general government department.

\end{itemize}

To address these challenges, the proposed intelligent agent, \textit{Pubbie}, incorporates carefully designed workflows and strategic model fine-tuning or few-shot learning approaches. To overcome the input length constraints and enable timely response, we innovatively integrate an SQLite database into our agent and leverage LLMs to transform natural language inquiries into SQL queries. A comprehensive discussion of these approaches will be presented in the subsequent subsections.

\subsection{Overview of the Workflows}

\begin{figure*}[htbp]
  \centering
  \includegraphics[scale=0.5]{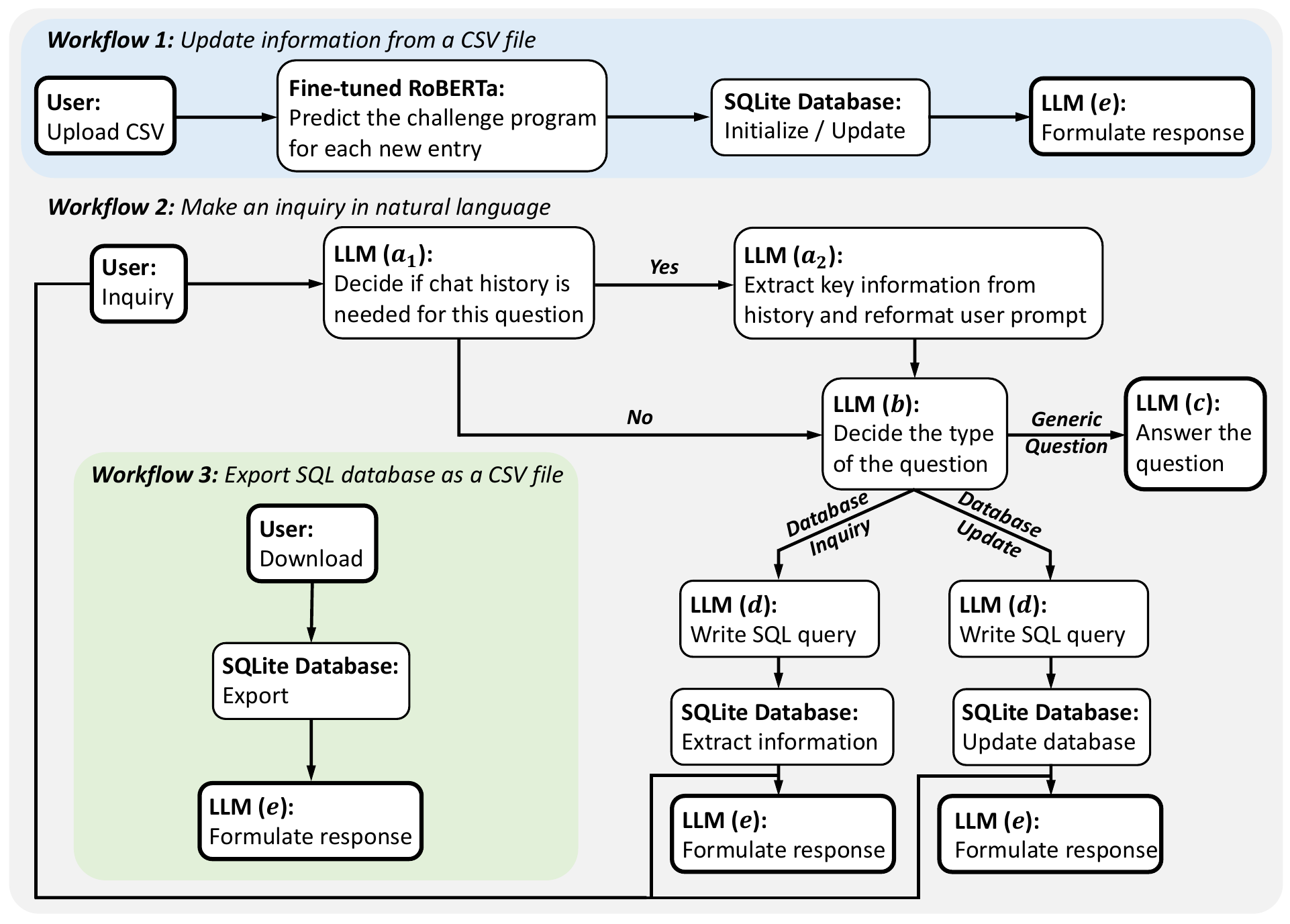}
  \caption{The three workflows of our intelligent agent, \textit{Pubbie}. The parenthesis after ``LLM'' indicates the type of prompt template utilized for few-shot learning.}
  \label{fig:workflow}
\end{figure*}

Based on its nature, our intelligent agent, \textit{Pubbie}, utilizes separate workflows to address the three primary tasks outlined in Section~\ref{sec:case_background}. As shown in Fig.~\ref{fig:workflow}, this involves LLM orchestration with few-shot learning via prompt templates, a fine-tuned RoBERTa model for semantic embedding \& program prediction, and an SQLite database for data storage \& management. 

\begin{itemize}
    \item	The \textbf{first workflow} (Fig.~\ref{fig:workflow} upper, in blue) handles the creation of the SQLite database. After the user uploads the CSV file by pressing the ``Browse files'' button (Fig.~\ref{fig:interface} middle right, circled in green), 10 selected attributes are fed into the fine-tuned RoBERTa model to automatically predict the corresponding challenge program for each publication\footnote{There are 12 challenge program categories and an additional ``no program'' category. If there is a ground-truth label of challenge program in the CSV file, this inference step will be omitted.}. The predicted challenge program and all attributes are then stored in the SQLite database. Once this process is finished, LLM (e) will respond to the user with a completion confirmation and a summary.
    \item The \textbf{second workflow} (Fig.~\ref{fig:workflow} middle, in grey) addresses the user’s inquiry (Fig.~\ref{fig:interface} bottom, circled in red) about NRC publications and challenge programs via LLM orchestration. To enrich responses with contextual information, \textit{Pubbie} employs LLM (a$_1$) to assess the relevance of chat history and then reformats the user’s prompt as needed using LLM (a$_2$).  Subsequently, LLM (b) categorizes the prompt as either a generic question, a database inquiry, or a database update. Generic questions are processed by LLM (c). When user prompts involve the backend SQLite database, \textit{Pubbie} employs an additional LLM (d) to transform the prompt into SQL queries, facilitating either data extraction or database updates\footnote{This enables the users to manually correct the predicted challenge program.}. Following this, the ultimate response is formulated by the subsequent LLM (e), which considers the user's prompt to structure the answer appropriately within the given context. The dual capability of independently addressing generic and database inquiries equips \textit{Pubbie} to tackle open-ended questions while providing accurate, in-depth contextual analysis.
    \item The \textbf{third workflow} (Fig.~\ref{fig:workflow} lower, in green) streamlines the process of exporting publications from the SQLite database into a downloadable CSV file. Upon retrieving the desired publications, users can initiate this process by clicking the ``Download CSV'' button (Fig.~\ref{fig:interface} lower left, circled in blue). Subsequently, pertinent data will be transcribed into a CSV file, with LLM (e) providing a summary. A pop-up window will then prompt the user to designate a storage location for the CSV file.
\end{itemize}

\begin{table*}[htbp]
    \centering
    \setlength{\tabcolsep}{3pt} 
    \fontsize{8}{10}\selectfont 
    \caption{Data flow from the user prompt to the agent response. The context-aware replacement is highlighted in bold.}
    \begin{tabular}{p{2cm}|p{4cm}|p{0.5cm}|p{4.5cm}|p{1cm}|p{4cm}}
         \hline\hline
        \multicolumn{1}{c|}{\textbf{\tabincell{c}{User\\Prompt}}} & 
        \multicolumn{1}{c|}{\textbf{\tabincell{c}{Prompt After\\Replacement}}} & 
        \multicolumn{1}{c|}{\textbf{\tabincell{c}{Question\\Type}}} & 
        \multicolumn{1}{c|}{\textbf{\tabincell{c}{SQL\\Query}}} &
        \multicolumn{1}{c|}{\textbf{\tabincell{c}{SQL\\Output}}} &
        \multicolumn{1}{c}{\textbf{\tabincell{c}{Agent\\Output}}} \\ 
        \hline\hline
        Hi! & \underline{LLM (a$_1$):} Hi! & \underline{LLM (b):} Generic & N/A & N/A & \underline{LLM (c):} Hello! How can I assist you today?  \\ 
        \hline
        What is \textbf{the data} about? & \underline{LLM (a$_2$):} What is \textbf{the NRC publication dataset} about? & \underline{LLM (b):} Generic & N/A & N/A & \underline{LLM (c):}  \textbf{The NRC publication dataset} contains information about publications at the National Research Council of Canada. It includes details such as titles ... \\ 
        \hline
        Give me the challenge program of \textbf{this publication}. & \underline{LLM (a$_2$):} Give me the challenge program of \textbf{the publication ``A Study on the Use of Intake Flow Path Modification to Reduce Methane Slip of a Natural Gas-Diesel Dual-Fuel Engine''}. & \underline{LLM (b):} SQL & \underline{LLM (d):} \texttt{SELECT prog FROM pub WHERE title = 'A Study on the Use of Intake Flow Path Modification to Reduce Methane Slip of a Natural Gas-Diesel Dual-Fuel Engine';} & Materials for Clean Fuels & \underline{LLM (e):} The challenge program for \textbf{the publication ``A Study on the Use of Intake Flow Path Modification to Reduce Methane Slip of a Natural Gas-Diesel Dual-Fuel Engine''} is ``Materials for Clean Fuels''.  \\ 
        \hline
        About \textbf{this author}, how many publications were written? & \underline{LLM (a$_2$):} About \textbf{Alysa}, how many publications were written? & \underline{LLM (b):} SQL & \underline{LLM (d):} \texttt{SELECT COUNT(*) FROM pub WHERE authors\_with\_affil LIKE '\%Alysa\%';} & 0 & \underline{LLM (e):} \textbf{Alysa} has not written any publications.  \\ 
         \hline\hline
    \end{tabular}
    \label{tab:sql}
\end{table*}

To deliver precise in-domain responses, few-shot learning is performed on all LLMs in Fig.~\ref{fig:workflow} utilizing prompt templates or domain-specific data. This will be explained in detail in the following Section III-D.

\subsection{Implementation Details}
The three-fold workflow presented in Fig.~\ref{fig:workflow} is implemented using the Semantic Kernel~\cite{microsoft_semantic_kernel} framework powered by Azure. Implementation details associated with the key components will be discussed in this subsection.

\subsubsection{Semantic embedding and classification} As most of the attributes of the publication are represented using plain text, it is essential to extract semantic embeddings before training a classification model. To this end, we compare two widely used models, i.e., BERT~\cite{devlin2019bert} and RoBERTa~\cite{zhuang2021obustly}. Ten semantically meaningful attributes, including title, author, affiliations, and keywords, are chosen as the input features, each mapped to a 768-dimensional embedding vector. 

Considering the limitation on computing resources\footnote{The training is conducted on a laptop with 32GB of RAM and an 8-core CPU.}, one linear layer is added on top of the pre-trained BERT\footnote{\url{https://huggingface.co/google-bert/bert-base-cased}} and RoBERTa\footnote{\url{https://huggingface.co/FacebookAI/roberta-base}} models during fine-tuning to classify the challenge program. Among the 656 labeled publications, 524 are allocated for fine-tuning, 66 for validation, and 66 for testing. The fine-tuned BERT model gives an accuracy of 81.8\% on the test set, while the fine-tuned RoBERTa model attains an accuracy of 84.8\%. Therefore, we proceed with integrating the fine-tuned RoBERTa model into our intelligent agent \textit{Pubbie}. A more comprehensive comparison with other widely adopted models is presented in the following Section~\ref{sec:eval}.

\subsubsection{LLM with few-shot learning} As out-of-the-box LLMs lack domain knowledge about the NRC's challenge programs, we conduct few-shot learning on GPT-3.5-Turbo\footnote{\url{https://learn.microsoft.com/en-us/azure/ai-services/openai/concepts/models\#gpt-35}} using prompt templates and domain data. As depicted in Figure~\ref{fig:workflow}, prompt templates for chat history detection \& assembly, question type classification, generic question handling, text-to-SQL conversion, and context-aware response formulation are respectively fed to LLM (a$_1$)-(a$_2$), LLM (b), LLM (c), LLM (d), and LLM (e). For instance, the prompt template on history replacement empowers the intelligent agent \textit{Pubbie} to disambiguate pronouns and other vague terms in user queries by cross-referencing ongoing queries with past chat history. Examples illustrating the data flow from the user prompt to the agent response are demonstrated in Table~\ref{tab:sql}.

To facilitate more accurate responses to generic questions using in-domain knowledge, we leverage the ``Bring your own data'' functionality\footnote{\url{https://techcommunity.microsoft.com/t5/educator-developer-blog/bring-your-own-data-to-azure-openai-step-by-step-guide/ba-p/3905212}} and index the key attributes of the publication dataset using Azure AI search service~\cite{MicrosoftAzureOpenAI}. Such data will empower the LLM responsible for generic questions (LLM (c)) to provide more precise answers. 

\subsubsection{Integration of the SQLite database} Due to the limitations on the input length of LLMs\footnote{\url{https://learn.microsoft.com/en-us/azure/ai-services/openai/concepts/models#gpt-35}}, feeding the entire publication dataset to the LLM all at once is impractical, rendering it impossible to correctly respond to inquiries that demand aggregated information. To address this challenge, a backend SQLite database is integrated into the intelligent agent to store both the publication dataset and the predicted challenge program. Such integration offers a dual advantage: 1) Given limited computational resources, loading the fine-tuned RoBERTa model and performing inference is time-consuming \cite{zhuang2021obustly}. By pre-storing the predicted program, response latency is significantly reduced. 2) Users are empowered to manually correct the predicted program using simple natural language queries. Any errors need to be rectified only once. Moreover, the serverless design of the SQLite database makes it an ideal option for managing confidential information in government~\cite{lloyd2024canada}.

\subsubsection{User-friendly interface design}
To enhance accessibility for non-technical users and lower the usability barriers, an intuitive user interface (UI) is incorporated into our intelligent agent utilizing Streamlit~\cite{khorasani2022web}, as illustrated in Fig.~\ref{fig:interface}. The primary mode of interaction between the user and the agent is typing queries in plain text in the chat box, while file uploads and downloads are operated by a simple click on buttons with clear visual cues (``Browse files'' or ``Download CSV'').

\subsection{System Evaluation}
\label{sec:eval}
This subsection presents the evaluation of the two primary components of the intelligent agent \textit{Pubbie}, i.e., the challenge program classification model and the text-to-SQL conversion mechanism.

\subsubsection{Challenge program classification} 

To explore the suitable choice of the classification model, we systematically evaluate the performance of the following four models: 1) fine-tuned BERT; 2) fine-tuned RoBERTa; 3) bag-of-words with a Naïve Bayes classifier~\cite{murphy2006naive}; and 4) out-of-the-box GPT-3.5-Turbo. The training, validation and test sets respectively comprise 524, 66 and 66 NRC publications with manually labeled challenge programs. The accuracy (acc.), precision (prec.), recall (rec.) and F1 score of these models on the test set are presented in Table~\ref{tab:result}.\footnote{The precision and recall are averaged at the program level.}

\begin{table}[h]
    \centering
    \caption{Performance on classifying the test set containing 66 NRC publications.}
    \setlength{\tabcolsep}{9pt} 
    \fontsize{10}{13}\selectfont 
    \begin{tabular}{c|c|c|c|c}
    \hline\hline
    \textbf{Model} & \textbf{Acc.} & \textbf{Prec.} & \textbf{Rec.} & \textbf{F1 Score} \\
    \hline\hline
    BERT & 0.818 & 0.606 & 0.612 & 0.588 \\
    \hline
    \textbf{RoBERTa} & \textbf{0.848} & \textbf{0.706} & \textbf{0.717} & \textbf{0.695} \\
    \hline
    Bag-of-Words & 0.621 & 0.362 & 0.370 & 0.349 \\
    \hline
    GPT-3.5-Turbo & 0.545 & 0.479 & 0.484 & 0.452 \\
    \hline\hline
    \end{tabular}
    \label{tab:result}
\end{table}

Based on Table~\ref{tab:result}, the fine-tuned RoBERTa obtains the best performance in terms of all evaluation metrics. Nonetheless, it is crucial to acknowledge that the model's performance is constrained by the limited size of the training data. Manual corrections can be performed easily by prompting \textit{Pubbie} with plain text, where the updates are documented and integrated into the back-end SQLite database.

\subsubsection{Text-to-SQL conversion} Leveraging LLMs to translate inquiries in plain text to SQL queries~\cite{shi2024survey} is proven to be more accurate than conventional deep learning approaches~\cite{kumar2022deep}. To evaluate the performance of the text-to-SQL conversion within our agent  (i.e., LLM (d)), we collect 26 inquiries related to the NRC's publication and the associated challenge program, comprising 13 frequently and 13 less frequently asked questions. Through a manual evaluation of the intermediary SQL queries generated by \textit{LLM (d)}, we observe that 25 of the queries are accurate (12 of 13 for frequently asked and 13 of 13 for less frequently asked questions), resulting in an overall accuracy of \textbf{96.15\%}.

In case \textit{Pubbie} encounters a SQL generation error, it notifies the user of the failure at runtime and suggests initiating a new prompt, allowing users to refine their queries on the fly.

\subsection{Cost Analysis}
\label{sec：cost}
\noindent This project leveraged existing NRC resources, including IT infrastructure and support systems. In addition, technical and subject matter guidance is provided by two NRC employees, each dedicating approximately 10\% of their time over a 4-month period to support the project. Incremental resources include the hiring of a second-year undergraduate computer science student for a 4-month term and the use of a dedicated environment in Azure OpenAI Studio, with the total incremental cost amounting to approximately \$18,000. Our solution is compatible with standard computers or laptops with a minimum of 8GB of RAM.

\subsection{Insights and Takeaways}
\label{sec:insights}
Reflecting on the development of the intelligent agent \textit{Pubbie}, there are several takeaways:

\subsubsection{Adapting to general government applications}

While the intelligent agent \textit{Pubbie} is tailored to address inquiries concerning the NRC's publications and associated challenge programs, the proposed workflows can be readily adapted to general governmental tasks that involve repetitive manual processes. By integrating backend LLMs, conventional AI models, and SQLite databases with a user-friendly front-end interface, this design can be applied to larger and more complex operational datasets.

\subsubsection{Capacity-building with non-technical users}
The user-friendly interface design of \textit{Pubbie} facilitates concise and transparent interactions between users and the intelligent agent. By enabling actions such as typing in the chat box or utilizing buttons for file upload/download, non-technical users are empowered to effectively engage with tools tailored to their requirements to minimize manual workloads. Moreover, the development of this intelligent agent is cost-effective in terms of computing resources and time of technical expertise, making it affordable for a general department in the government. 

\subsubsection{Collective problem-solving}
The case study exemplifies one of the 22 projects launched by the NRC's GenAI initiative, where each project is assigned to one student co-supervised by a business process owner and a computer science researcher. Apart from weekly meetings organized by individual projects, a collective weekly meeting is organized for all 22 students to conduct brainstorms and share ideas. This collaborative problem-solving approach values diverse stakeholder perspectives and fosters a collective effort to address common challenges across projects.

\section{Conclusion}
\label{sec:conclusion}

This paper presents the pioneering exploration of utilizing Generative AI techniques to facilitate government operations. Among the 22 projects from the GenAI initiative at the National Research Council of Canada (NRC), we focus on the development of the intelligent agent \textit{Pubbie}, targeting streamlining and automating the data management and reporting workflows at the NRC. State-of-the-art techniques such as LLM orchestration and semantic embedding via RoBERTa are incorporated with carefully designed fine-tuning and few-shot learning strategies to infuse domain knowledge for accurate response. Our system design accounts for constraints on computing resources, while the inclusion of a user-friendly interface aims to reduce usability barriers. Applying the intelligent agent \textit{Pubbie} in the daily work at the NRC’s performance measurement team greatly reduces the reliance on manual efforts and improves operation efficiency. Upon the establishment and deployment of Pubbie within the NRC, existing processes can be streamlined, leading to efficiency gains throughout the organization and increased engagement with the digital solution. Future work will focus on improving the stability of the intelligent agent, extending the design to more usage scenarios, and encompassing other performance measurement datasets.

\section{Acknowledgement}
\label{sec:acknowledgement}
\vspace{3pt}
The authors gratefully acknowledge the support of the National Research Council’s Chief Digital Research Officer \& Chief Science Officer, Dr. Joel Martin, and Generative AI Innovation Lead, Daniel Lowcay, whose vision and leadership in establishing the NRC’s AI Innovation Initiative for Operational Excellence, along with their invaluable advice, were instrumental to this project’s success. The authors also thank Dr. Patrick Littell \& Dr. Rebecca Knowles from the NRC’s Multilingual Text Processing team for their support and review of this paper. Finally, the essential contributions of the NRC’s Performance Measurement \& Accountability Reporting team, who generously provided their time and creativity to support \textit{Pubbie}’s development and testing, are deeply appreciated. This project would not have been possible without their support.

\ifCLASSOPTIONcaptionsoff
  \newpage
\fi

\bibliographystyle{IEEEtran}
\bibliography{main}

\end{document}